\documentclass[aps,superscriptaddress,superbib,twocolumn,groupedaddress,showpacs]{revtex4}

\usepackage[highlight]{neel}

\bibpunct{[}{]}{,}{n}{}{}

\graphicspath{{fig/}}

\usepackage[latin1]{inputenc}
\usepackage[T1]{fontenc}

\def\figurename{Fig.}

\begin{document}

\renewcommand\topfraction{0.8}
\renewcommand\bottomfraction{0.7}
\renewcommand\floatpagefraction{0.7}

\title{Magnetic domain walls in nanostrips of single-crystalline $\mathrm{Fe}_4\mathrm{N}$(001) thin films with fourfold in-plane magnetic anisotropy}%

\author{Keita Ito}
\affiliation{Institute of Applied Physics, Graduate School of Pure and Applied Sciences, University of Tsukuba, Tsukuba, Ibaraki 305-8573, Japan}%
\affiliation{Department of Electronic Engineering, Graduate School of Engineering, Tohoku University,  Sendai 980-8579, Japan}%
\affiliation{Japan Society for the Promotion of Science (JSPS), Chiyoda, Tokyo 102-0083, Japan}%
\affiliation{Univ. Grenoble Alpes, Inst NEEL, F-38000 Grenoble, France}%
\affiliation{CNRS, Inst NEEL, F-38000 Grenoble, France}%
\affiliation{Tsukuba Nanotechnology Human Resource Development Program, Graduate School of Pure and Applied Sciences, University of Tsukuba, Tsukuba, Ibaraki 305-8571, Japan}

\author{Nicolas Rougemaille}
\affiliation{Univ. Grenoble Alpes, Inst NEEL, F-38000 Grenoble, France}%
\affiliation{CNRS, Inst NEEL, F-38000 Grenoble, France}

\author{Stefania Pizzini}
\affiliation{Univ. Grenoble Alpes, Inst NEEL, F-38000 Grenoble, France}%
\affiliation{CNRS, Inst NEEL, F-38000 Grenoble, France}

\author{Syuta Honda}
\affiliation{Institute of Applied Physics, Graduate School of Pure and Applied Sciences, University of Tsukuba, Tsukuba, Ibaraki 305-8573, Japan}%
\affiliation{Tsukuba Nanotechnology Human Resource Development Program, Graduate School of Pure and Applied Sciences, University of Tsukuba, Tsukuba, Ibaraki 305-8571, Japan}

\author{Norio Ota}
\affiliation{Tsukuba Nanotechnology Human Resource Development Program, Graduate School of Pure and Applied Sciences, University of Tsukuba, Tsukuba, Ibaraki 305-8571, Japan}

\author{Takashi Suemasu}
\affiliation{Institute of Applied Physics, Graduate School of Pure and Applied Sciences, University of Tsukuba, Tsukuba, Ibaraki 305-8573, Japan}%
\email[]{suemasu@bk.tsukuba.ac.jp}

\author{Olivier Fruchart}
\affiliation{Univ. Grenoble Alpes, Inst NEEL, F-38000 Grenoble, France}%
\affiliation{CNRS, Inst NEEL, F-38000 Grenoble, France}%
\email[]{Olivier.Fruchart@neel.cnrs.fr}

\date{\today}


\newcommand\FeN{\ifmmode{\mathrm{Fe}_4\mathrm{N}}\else{$\mathrm{Fe}_4\mathrm{N}$\ \xspace}\fi}

\begin{abstract}

We investigated head-to-head domain walls in nanostrips of epitaxial $\FeN(001)$ thin films, displaying a fourfold magnetic anisotropy. Magnetic force microscopy and micromagnetic simulations show that the domain walls have specific properties, compared to soft magnetic materials. In particular, strips aligned along a hard axis of magnetization are wrapped by partial flux-closure concertina domains below a critical width, while progressively transforming to zigzag walls for wider strips. Transverse walls are favored upon initial application of a magnetic field transverse to the strip, while transformation to a vortex walls is favored upon motion under a longitudinal magnetic field. In all cases the magnetization texture of such fourfold anisotropy domain walls exhibits narrow micro-domain walls, which may give rise to peculiar spin-transfer features.

\end{abstract}

\maketitle



\section{Introduction}

A magnetic domain wall~(DW) may be set in motion under the stimulus of an electric current\cite{bib-GRO2003b,bib-YAM2004b}. The first phenomena mentioned to allow this are so-called direct spin-torque effects, responsible for transfer of angular momentum and direct exchange from the spin of conduction electrons to magnetization\cite{bib-BER1996,bib-SLO1996}. Later, other sources of transfer of momentum and exchange torque were proposed such as Rashba and spin-Hall effects\cite{bib-MOO2008,bib-KIM2010a,bib-BAN2012,bib-RYU2013,bib-EMO2013}. Novel devices making use of movable DWs in strips have been proposed, such as logic gates, spin-torque oscillators, amplifiers and non-volatile storage cells\cite{bib-BAR2006,bib-FRA2008,bib-PAR2008,bib-LEP2010}. In the case of spin torques arising from conduction electrons flowing inside the ferromagnet, the efficiency of the charge current to drive DW motion is expected to depend on the polarization of conduction electrons $P_\sigma = (\sigma_\uparrow - \sigma_\downarrow)/(\sigma_\uparrow + \sigma_\downarrow)$, with $\sigma_i$ the conductivity of electrons of a given spin state. The sign of $P_\sigma$ should determine the direction of motion: along the electron flow for $P_\sigma>0$, and against it for $P_\sigma<0$.

Here we are considering the ferromagnetic material $\FeN$. It is an anti-perovskite metal nitride, with a negative $P_\sigma$ predicted theoretically ($P_\sigma=-1$)\cite{bib-KOK2006} and confirmed experimentally\cite{bib-KOK2012,bib-ITO2014}. The Curie temperature is reported to be $\tempK{767}$\cite{bib-SHI1962}, which eases fundamental investigations while letting the door open for possible applications. Therefore, \FeN is an interesting candidate material for investigating current-driven DW motion under negative $P_\sigma$. The spontaneous magnetization and cubic magnetic anisotropy energy density have been reported at room temperature in $\mathrm{Fe}_4\mathrm{N}$ epitaxial films grown on MgO(001) by molecular beam epitaxy (MBE)\cite{bib-COS2004}: $\Ms=\unit[1.43]{\mega\ampere\per\meter}$ and $\Kc=\unit[2.9{\times}10^4]{\joule\per\meter\cubed}$, which makes \FeN a reasonably-soft magnetic material. The in-plane <100> directions of \FeN films are easy magnetization directions, whereas the in-plane <110> directions are hard axes\cite{bib-COS2004}.

In this manuscript we report magnetic force microscopy~(MFM) and micromagnetic simulations of head-to-head DWs in nanostrips of $\FeN(001)$ epitaxial films. We first present methods and results\bracketsecref{sec-results}. We then analyze the MFM images to derive the distribution of magnetization in the strips\bracketsecref{sec-mfm-analysis}. We finally discuss the resulting distributions on the basis of analytical arguments and micromagnetic simulations\bracketsecref{sec-discussion}.

\section{Results}
\label{sec-results}

\begin{figure}
  \begin{center}
  \includegraphics[width=74.362mm]{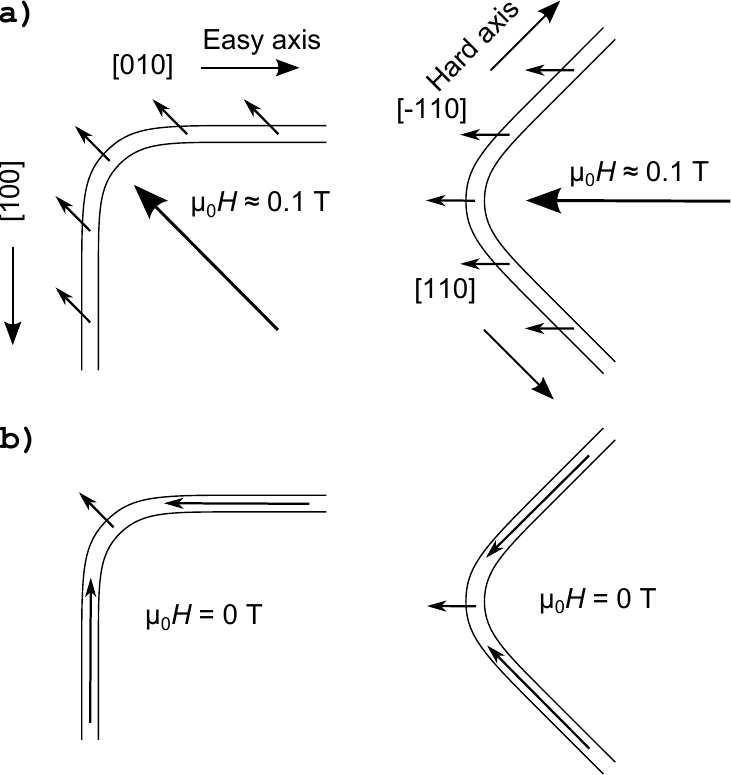}%
  \caption{\label{fig-geometry}{Geometry of the strips investigated}Curved \FeN wires with their arm directions set either along the <100> easy or <110> hard magnetization directions. (a)~Magnetization step before imaging (b)~DWs at remanence.}
  \end{center}
\end{figure}

We grew by MBE a $\mathrm{Au}(\lengthnm{3})/\FeN(\lengthnm{t\mathrm{=}10})$ stack on SrTiO$_3$(001)[19]. Epitaxial growth of the \FeN layer was confirmed by reflection high-energy electron diffraction and symmetric out-of-plane x-ray diffraction\cite{bib-ITO2014}. The purpose of Au is to prevent surface oxidation of the magnetic layer. The stack was patterned into strips with widths~$w$ ranging from $\lengthnm{200}$ to $\unit[2]{\micro\meter}$, using a combination of electron-beam lithography and ion-beam etching. The strips consist of two linear parts making an angle \angledeg{90} one with another, connected with a circular part with large radius. Two sets of strips have been patterned, with the linear parts aligned along either <100> or <110> axes, see \figref{fig-geometry}.

Following a standard procedure\cite{bib-TAN1999} we applied an in-plane magnetic field along the radius of the bends, with a strength $\approx\unit[0.1]{\tesla }$ exceeding that related to the transverse shape anisotropy. Upon reducing the field back to zero this gives rise to DWs at the bends, with the strip locally oriented along an easy or hard magnetization axis, depending on the set of patterns. Domain walls may be of either head-to-head or tail-to-tail nature depending on the sign of the applied field. Both are equivalent under time-reversal symmetry, so for the sake of simplicity we will use the name head-to-head to cover both types. The DWs were investigated using MFM with a NT-MDT Ntegra atomic force microscope operated under ambient conditions. We used low-moment tips to prevent dragging DWs during imaging. These are AC-240TS Asylum force modulation non-magnetic cantilevers and tips, sputter-coated by a $\mathrm{Al}_2\mathrm{O}_3(3)/\mathrm{Pt}(1)/\mathrm{Co}_{80}\mathrm{Cr}_{20}(10)$. The numbers in brackets stand for the thickness in nanometers, and the materials deposited first are mentioned to the right of the stack.  The typical radius of curvature and effective magnetic moment of the tips are \lengthnm{10} and $\unit[5\times10^{-18}]{\ampere\usk\meter\squared}$, respectively. Images were made in a two-pass scheme in modulation of amplitude during the first pass, and monitoring the phase during the second pass. The peak-to-peak amplitude of oscillation was \lengthnm{50}, and the lift height during second pass was set to a few nanometers. Negative~(resp. positive) contrast stands for attractive~(resp. repulsive) forces. The exact width of the strips was derived from the topographic images, inferred from the width measured at the top of the strip to minimize the apparent broadening due to the finite size of the tips.

Micromagnetic simulations were conducted using the OOMMF software\cite{bib-OOMMF-report,bib-OOMMF}. Material parameters are magnetization $\Ms=\unit[1.43]{\mega\ampere\per\meter}$, exchange stiffness $A=\unit[15]{\pico\joule\per\meter}$, cubic anisotropy~$\Kc=\unit[3\times10^{4}]{\joule\per\cubic\meter}$ or $\Kc=\unit[5\times10^{4}]{\joule\per\cubic\meter}$. Damping is set to $\alpha=1$ to speed up convergence, with no consequence on the result as far as equilibrium states are concerned. The cell size is $\unit[4\times4\times10]{\nano\meter\cubed}$, chosen so that reducing it further yields only negligible differences. Magnetic moments at the two extremities of the strips are fixed to avoid non-uniform magnetization distributions at the edges. The length-over-width ratio of the strips is chosen equal at least to ten, to limit finite-size effects. We use the volume density of magnetic charges $-\mathrm{div}\vect M$ as an indication for the MFM contrast.

\begin{figure}
  \begin{center}
  \includegraphics[width=84.313mm]{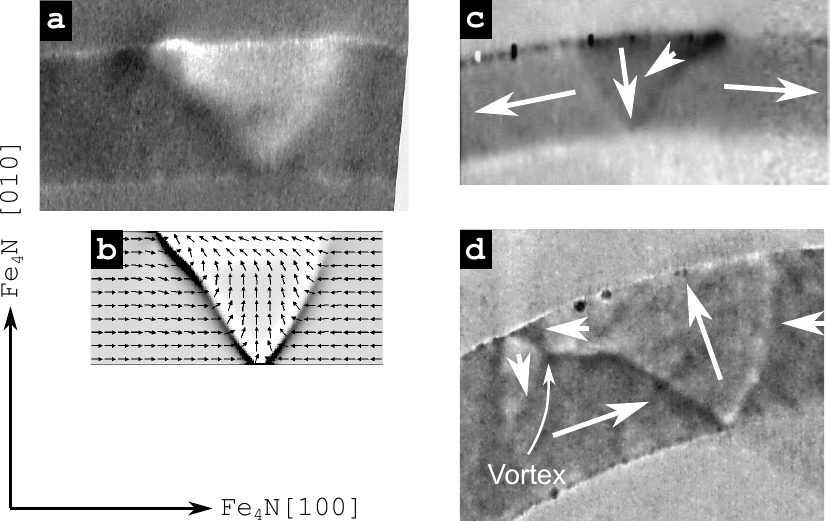}%
  \caption{\label{fig-long-anisotropy}DWs as a function of strip width for sides locally parallel to [001], an easy magnetocrystalline axis. The strip width is (a,b)~\lengthnm{320}, (c)~\lengthnm{520} and (d)~\lengthnm{1080}. In (a,c,d) the DWs are imaged with MFM, and are of either head-to-head~(a,d) or tail-to-tail~(c) DWs type. (b)~is the map of magnetic charges for a simulation based on anisotropy $\Kc=\unit[\scientific{3}{4}]{\joule\per\cubic\metre}$.  Arrows depict the local direction of magnetization. }
  \end{center}
\end{figure}

\begin{figure}
  \begin{center}
  \includegraphics[width=87.864mm]{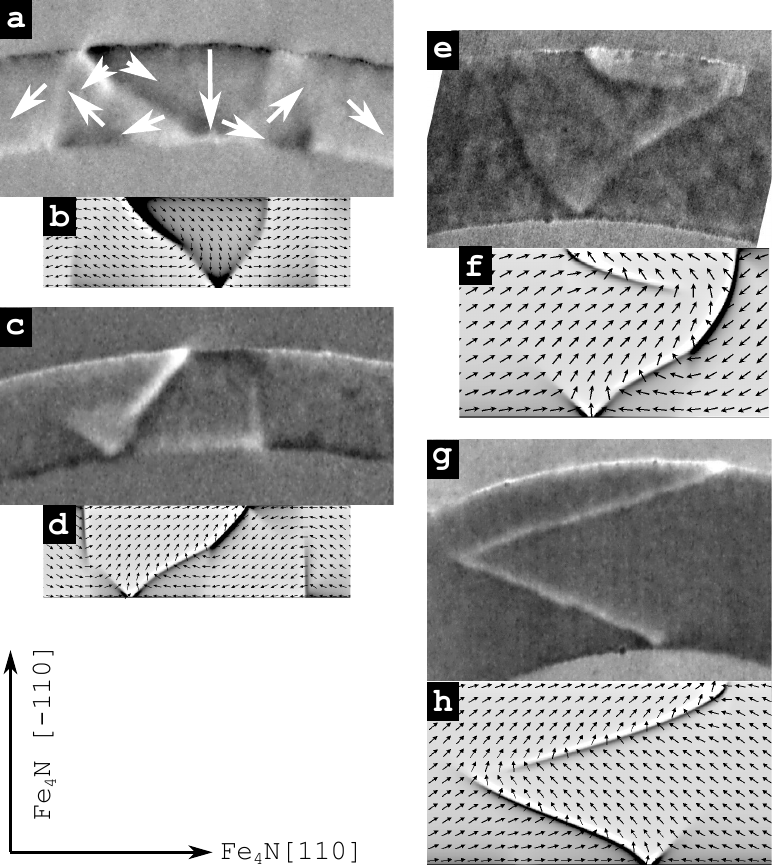}%
  \caption{\label{fig-tilted-anisotropy}DWs as a function of strip width for side locally parallel to $\mathrm{[1-10]}$, a hard magnetocrystalline axis. The strip width is (a,b)~\lengthnm{320}, (c,d)~\lengthnm{600}, (e,f)~\lengthnm{800} and (g,h)~\lengthnm{1100}. The DWs are of either head-to-head~(c-h) or tail-to-tail~(a,b) type. For simulations the map is that of magnetic charges for simulations based on anisotropy $\Kc=\unit[\scientific{3}{4}]{\joule\per\cubic\metre}$. (b,d)~are configurations at rest, while (f,h)~are a transient state, evolving very slowly towards a configuration similar (c). Arrows depict the local direction of magnetization.}
  \end{center}
\end{figure}

\figref{fig-long-anisotropy} shows experimental and simulated MFM patterns of DWs in strips locally parallel to the magnetic easy directions of the material. \figref{fig-tilted-anisotropy} is the analogous, with the strip sides locally parallel to a magnetic hard direction of the material. In \figref{fig-after-motion}, DWs have been moved from their initial location, with the use of a magnetic field of a few mT. The origin of the MFM contrast and the derivation of the associated magnetic patterns is discussed in the next section.

\section{MFM contrast and micromagnetic simulations}
\label{sec-mfm-analysis}

The present MFM tips with low moment and magnetic material essentially at the apex, provide a fair picture of the second derivative of the stray field emanating from the sample\cite{bib-KRA1995b}, with an only weak contribution of the always-attractive and so-called susceptibility contrast\cite{bib-HUB1997}. As the tips are magnetized mainly along the normal to the sample, we shall discuss this contrast as qualitatively representative of the vertical stray field arising from the sample. The tips are magnetized downwards, so that positive~(resp. negative) contrast highlights upwards~(resp. downwards) stray field, thus arising from positive~(resp. negative) charges.

\begin{figure}
  \begin{center}
  \includegraphics[width=82.677mm]{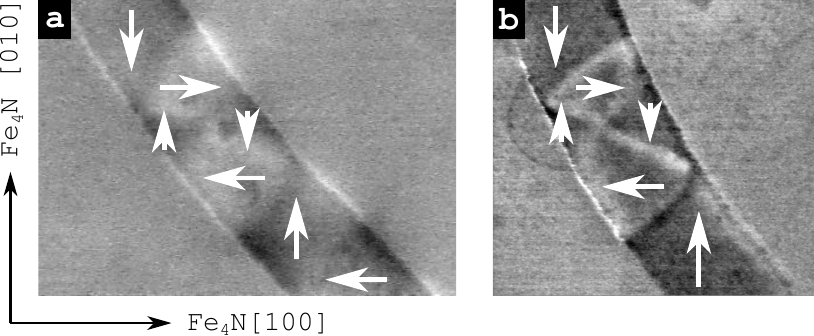}%
  \caption{\label{fig-after-motion}Head-to-head DWs after motion along strips under a magnetic field of a few mT. At the initial location of the DWs~(bend not displayed here, located to the upper left side of the images) the strip was locally parallel to an easy axis, while at the final location it is close to a hard axis direction. The strip width is (a)~\lengthnm{410} and (b)~\lengthnm{550}.}
  \end{center}
\end{figure}

Stray fields arising from in-plane magnetized samples may give rise to various types of patterns. At strip side edges, positive~(resp. negative) contrast arises for magnetization pointing towards (resp. away from) a side edge, reflecting surface magnetic charges~$\vect M\dotproduct\vect n$, while magnetization parallel to an edge yields no contrast. Away from edges, stray fields may arise from volume charges~$-\mathrm{div}\vect M$ related to non-homogeneous magnetization patterns such as DWs. DWs are expected to be of N\'{e}el type for the thickness considered, \ie magnetization rotates within the plane inside the DW\cite{bib-HUB1998b}. In extended thin films, N\'{e}el walls tend to be magnetically uncharged as a whole, to minimize long-range magnetostatic energy. To do so, unless some specific constraints are applied, a DW bisects the direction of magnetization in the neighboring domains, so that magnetic charges are balanced. This gives rise to a balanced bipolar contrast, of strength increasing with the wall angle\cite{bib-HUB1999}. However in strips DWs are of head-to-head type, and carry the total charge~$2tw\Ms$. Thus, a net monopolar contrast is expected over the global pattern. These reminders allow us to determine the distribution of magnetization from the MFM contrast, and compare them with micromagnetic simulations~(FIGs.\ref{fig-long-anisotropy}-\ref{fig-after-motion}).

When the easy directions of magnetization are parallel and transverse to the strip\bracketfigref{fig-long-anisotropy}, the arrangement of magnetization is very similar to transverse walls in soft magnetic materials, in which the monopolar charge is mostly uniformly distributed over its triangular shape\cite{bib-CHA2010}. The area of the wall scales with $w^2$ while the total charge scales with~$w$. Thus the monopolar contrast decreases with increasing strip width, making the dipolar contrast on the $90\deg$ N\'{e}el walls larger in relative values, thus globally more visible\bracketsubfigref{fig-long-anisotropy}d.

The situation is more complex when the hard directions of magnetization are parallel and transverse to the strip\bracketfigref{fig-tilted-anisotropy}. For narrow width the core of the DW has a shape similar to that of an asymmetric transverse wall\cite{bib-NAK2005}, and bears most of the monopolar contrast\bracketsubfigref{fig-tilted-anisotropy}a. However, on either side of the DW, alternating light and dark contrast is seen on the strip side edge, revealing alternating sub-domains with magnetization with angles $\pm\pi/4$, in a concertina fashion\cite{bib-STE2011}. Between these sub-domains the light/dark contrast over the strip hints at a narrow N\'{e}el wall pinched at only one edge, a pattern which may be viewed as an edge half-vortex\cite{bib-TCH2005}. Upon increasing the strip width the monopolar contrast decreases, similar to the case of~\figref{fig-long-anisotropy}. The tendency to form side sub-domains progressively disappears~\bracketsubfigref{fig-tilted-anisotropy}{c-f}. For width above typically one micrometer the N\'{e}el wall outlining the triangular shape of the transverse wall breaks into two or more segments, yielding a zigzag shape\bracketsubfigref{fig-tilted-anisotropy}{g-h}. The dominant contrast is monopolar, superimposed with weaker dipolar contrast. This combination is consistent with the observation that the direction of the DW does not bisect the direction of magnetization in the neighboring domains. These micro-DWs contain most of the total head-to-head charge $2t w \Ms$ of the entire wall.

In \figref{fig-after-motion} the domains are clearly of the vortex type. As in the previous two figures the areal density of magnetic charges is larger for narrower strips, so that in \subfigref{fig-after-motion}a it is mostly homogeneous over the triangular areas of the DW as for soft magnetic materials, whereas DW contrast is more visible in~\subfigref{fig-after-motion}b.

\section{Discussion}
\label{sec-discussion}

In this section we first provide analytical arguments pertaining to strip with a fourfold magnetic anisotropy, and then discuss the experimental observations as well as the micromagnetic results.

\subsection{Characteristic length and energy scales}
\label{sec-discussionScales}

Let us first derive characteristic micromagnetic quantities of $\mathrm{Fe}_4\mathrm{N}$ based on the known magnetic properties of the material. In a thin film, magnetization is essentially in-plane and can thus be described by a unique in-plane angle $\theta$ with reference to \eg [100]. The volume density of magnetocrystalline anisotropy will be described to first order with $\Emc=(\Kc/4)\sin^2(2\theta)$, with $\Kc\approx\unit[2.9\times10^4]{\joule\per\cubic\meter}$\cite{bib-COS2004}. The Curie temperature of \FeN is $\unit[\approx25]{\%}$ smaller than that of Fe, so that to first approximation we will use for exchange stiffness $A=\unit[1.5\times10^{-11}]{\joule\per\meter}$. Magnetization is such that $\muZero\Ms=\unit[1.8]{\tesla}$. Let us discuss the length and energy scales resulting from these numbers. Easy directions of magnetization in the cubic material \FeN are orthogonal one to another, so that it is natural to consider $90\deg$ DWs. In the absence of dipolar energy the width of such a DW would be $(\pi/2)\sqrt{4A/\Kc}=\pi\CubicAnisotropyExchangeLength\approx\lengthnm{70}$, and its energy per unit length  $t\sqrt{A\Kc}\approx\unit[6.7]{\pico\joule\per\meter}$. Another characteristic length scale is the dipolar exchange length $\DipolarExchangeLength$ such that $\pi\DipolarExchangeLength=\pi\sqrt{A/\Kd}\approx\lengthnm{10}$, with $\Kd=\frac12\muZero\Ms^2\approx\unit[1.43\times10^{6}]{\joule\per\cubic\meter}$ the dipolar constant. $\DipolarExchangeLength$ is typically the width of dipolar-charged N\'{e}el walls, with energy per unit length $4t\DipolarExchangeLength\Kd=4t\sqrt{A\Kd}\approx\unit[185]{\pico\joule\per\meter}$\cite{bib-HUB1998b}. From these figures it is clear that the dipolar energy will be the leading term to determine the shape of the domain walls, while the magnetocrystalline anisotropy may have an impact only in those situation where the former is weak due to the geometry.

\subsection{Domain walls for strips parallel to an easy axis}
\label{sec-discussionEasyAxis}

The case of DWs in strips locally aligned along an easy axis, is the easiest to discuss. In transverse walls in a soft magnetic material, the direction of magnetization is mostly along either the longitudinal or transverse direction of the strip, at least for small to moderate strip width\cite{bib-MIC1997,bib-NAK2005}. Thus, as these directions are easy axes for the fourfold anisotropy material in this first case, no significant extra cost is expected. It is therefore understandable that transverse walls are observed as a stable state\bracketfigref{fig-long-anisotropy}. For larger width an asymmetry appears markedly, as already observed for strips made of a soft magnetic material\cite{bib-CHA2010}. The observed increase of asymmetry with width is consistent with simulations made for soft magnetic materials\cite{bib-NAK2005}. Our micromagnetic simulations taking into account $\Kc$, reproduce faithfully the shape of these walls\bracketsubfigref{fig-long-anisotropy}a.

However, like for the case of soft magnetic materials, simulations also point at the existence of vortex walls, which are of lower energy for the strip widths considered here. The absence of the ground-state vortex walls in the experiments is however not surprising, similar to the well-established case of soft magnetic materials: transverse walls are favored by the preparation procedure under a transverse field. Only for strip thickness above typically \lengthnm{15} for materials with large magnetization such as Co\cite{bib-KLAe2004}, would vortex walls appear spontaneously. For the present samples, the only case where we observed vortex walls, is upon DW motion under an applied field\bracketfigref{fig-after-motion}. Note however that in the final state the strips are no longer parallel to the easy axis. Thus, the DWs in \figref{fig-after-motion} should be compared to those in \figref{fig-tilted-anisotropy} discussed below, where the strips are parallel to the hard axis, however again mostly of transverse type.

The transformation from transverse to vortex wall upon motion is analogous to the case of strips made of a soft magnetic film. This reflects the expected transformation of DWs from transverse to vortex or antivortex during motion above the Walker field\cite{bib-THI2006}. This is also consistent with the expected lower energy of vortex walls against transverse walls for our material parameters and strip geometries.

\subsection{Domain walls for strips parallel to a hard axis}
\label{sec-discussionHardAxis}

The case of DWs in strips with sides locally along hard axes requires more discussion. From the MFM contrast inferred and the simulation presented in \secref{sec-mfm-analysis}, it is clear that the local direction of magnetization is essentially along an easy direction of magnetization, \ie tilted with a $\pm\pi/4$ angle. A uniform state with magnetization aligned parallel to the strip side would allow a gain of dipolar energy with average volume density $(1/\sqrt{2})^2 N_\mathrm{t}\Kd\approx(t/w)\Kd(\ln2)/2$ with $N_\mathrm{t}$ the transverse demagnetizing factor~(the $\approx(1/\sqrt{2})^2$ pre-factor is related to the $\pi/4$ angle between magnetization and the direction of strip side). However, as this would compete with a cost $\Kc/4$ in magnetocrystalline anisotropy, one expects the following length scale to show up: $w\approx2t(\Kd/\Kc)\ln2\approx\lengthnm{300}$. Experimentally, is is indeed found that magnetization is nearly parallel to the strip direction for $w=\unit[200]{\nano\metre}$, and strictly parallel for narrower strips. To understand what happens for larger widths, let us recall that dipolar fields are short-ranged in 2D systems\cite{bib-FRU2005e}, being strong close to the side edges and weak at the center of structures. Thus, in most of the strip the demagnetizing field is much weaker than the demagnetizing coefficient would suggest. As long as the width is significantly larger than $\CubicAnisotropyExchangeLength\approx\unit[25]{\nano\meter}$, which is the case here, magnetization is expected to be essentially aligned with the side at a distance below $\CubicAnisotropyExchangeLength$, while recovering a direction close to an easy magnetocrystalline direction at larger distances. This is clearly confirmed by simulations~(see for example \figref{fig-tilted-anisotropy}. Besides, the occurrence of alternating side domains around the DW allows to further decrease dipolar energy compared to the nearly uniformly-magnetized state along a $\pi/4$ direction compared to the strip side direction. This is similar to Landau domains in films with a moderate perpendicular magnetocrystalline anisotropy, also called stripe domains\cite{bib-LAN1935}. Such patterns have already been reported as the ground state in strips made of an epitaxial material with an easy direction of magnetization across the strip, eg with Fe(110)\cite{bib-RUE1998} or Co(10-10)\cite{bib-PRE2001}. Consistently, even without DWs, we observed stand-alone stripe domains for width larger than $\unit[200]{\nano\metre}$~(inset of \figref{fig-concertina}). The existence of stripe domains is confirmed by simulations, based on magnetic parameters known a priori for our material\bracketfigref{fig-concertina}. The situation of alternating domains was modeled by Kittel in thin films with perpendicular magnetization, predicting the persistence of domains for thick films, with a period scaling like~$\sqrt{t}$\cite{bib-KIT1949b}. In such thin films, the period increases with thickness as a compromise between the increasing cost of domain walls extending along the thickness, and the gain in magnetostatic energy, arising from charges mostly at surfaces and with a certain range. Coming back to the present case, the width of the strip is analogous to the film thickness in Kittel's model, and the total energy of a domain wall increases with the strip width. However, the closure of flux arising from side edge charges becomes quite ineffective at long distances due to the short range of dipolar interactions in one-dimensional systems. This is confirmed both by experiments and by simulations, with the signature that the period of the stripes rises sharply with the strip width. The stripe domains becomes even unfavorable for any period, for too large strip width, typically $\unit[1]{\micro\meter}$.

As presented in \secref{sec-mfm-analysis}, for very large strips DWs consist mostly in a tilted line of charges across the strip. For strip width above typically $\unit[1]{\micro\meter}$, experimentally these lines often break into several segments with a zigzag shape\bracketsubfigref{fig-tilted-anisotropy}{g-h}. Head-to-head zigzag DWs are known in extended thin films with uniaxial magnetocrystalline anisotropy\cite{bib-HUB1979,bib-HUB1981}. They occur because they allow to spread the magnetic charges and thus decrease magnetostatic energy. Their angle results from the balance between reduction of magnetostatic energy and increase of DW length. A second-order effect is that the zigzag slightly decreases the long-range magnetostatic energy compared to a linear DW. Zigzag DWs in strips in a material with cubic anisotropy had already been reported\cite{bib-POK1997,bib-FON2011}, although many details had not been outlined, such as the distribution of charges, the possible side concertina and the cross-over from an asymmetric transverse DW to a zigzag DW.

\subsection{Quantitative analysis versus material imperfections}
\label{sec-discussionQuantitative}

\begin{figure}
  \begin{center}
  \includegraphics[width=77.814mm]{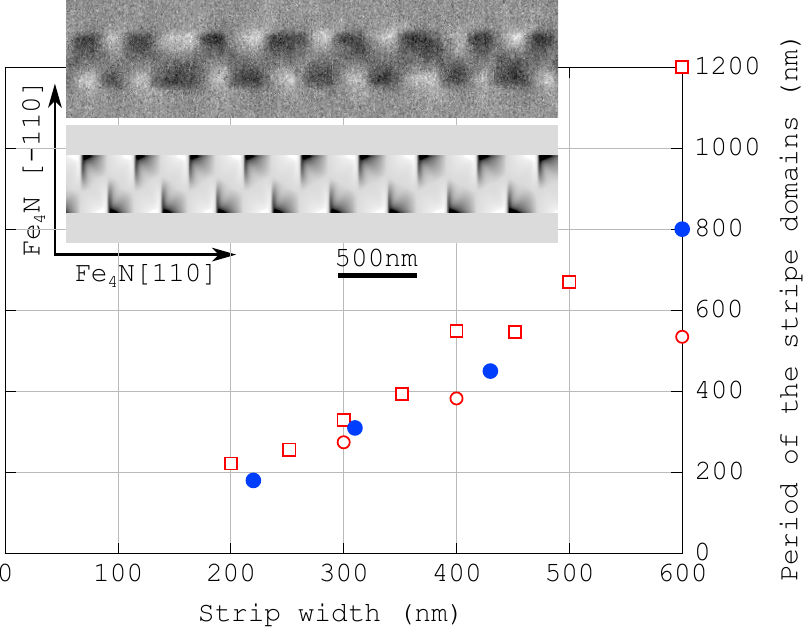}%
  \caption{\label{fig-concertina}Features of the stripe-domain pattern for strip with side parallel to a hard axis direction. Period (two domains) of the pattern versus the strip width. Open squares and disks are simulation results with magnetic anisotropy $\Kc$ $3\times10^4$ and $\unit[5\times10^4]{\joule\per\meter\cubed}$, respectively. Solid disks is the experimental lower bound for the local period. Inset: experimental~(phase contrat $\unit[0.10]{\deg}$) and simulated MFM contrast for the stripe phase~(strip width $\unit[420]{\nano\metre}$ and $\unit[400]{\nano\metre}$, respectively). }
  \end{center}
\end{figure}

The agreement of experiments and simulations is often striking~(\figref{fig-long-anisotropy} and \figref{fig-tilted-anisotropy}), raising the question of a possible quantitative analysis. In principle, one could extract some information on the material magnetic parameters (for example, $\Kc$), from features such as the width for which stripe domains disappear, or more generally on the width-dependence of their period\bracketfigref{fig-concertina}. In practice however, hysteresis effects related to material imperfections may prevent ground states to be reached. This is clear on \figref{fig-concertina}, where the stripe period suffers from fluctuations. Starting from a homogeneously-magnetized state at remanence, one may argue that the experimental domain size will be larger than the equilibrium size, due to hysteresis. Such effects are expected to gain in importance for rather large strips and very thin films as in the present case, because the energy landscape becomes very flat due to the short range of dipolar interactions.

Another hint of the interplay of a very flat energy landscape with material imperfections is provided by \subfigref{fig-tilted-anisotropy}{e-h}. While the agreement between simulations and experiments is excellent, it must be stressed that the simulations are not equilibrium states. These are snapshots taken during the evolution towards a simple asymmetric transverse wall. This evolution is very slow, which hints at a very flat energy landscape. It is likely that in the experiments this slow evolution was frozen due to pinning of the domain wall on local imperfections.

\section*{Conclusion}
\label{sec-conclusion}

In conclusion, we have studied the detailed inner structure of head-to-head domain walls in the negative and high-spin-polarization material $\mathrm{Fe}_4\mathrm{N}$, obtained epitaxially with the $(001)$ orientation. The in-plane fourfold anisotropy induces modifications in the distribution of magnetization in the walls, compared to soft magnetic materials. In particular, when the strip is locally parallel to a hard magnetization axis, below a critical strip width the domain walls are dressed with side structures reminiscent of concertina and stripe domain patterns. Above this width they progressively turn into sharp zig-zag domain walls, displaying an asymmetric charge and thus MFM contrast. In all cases the micro-domain walls in the structure are rather narrow, which may give rise to peculiar features of spin transfer torques\cite{bib-JAN2015}.

\section*{Acknowledgements}
\label{sec-thanks}

This work was financially supported in part by a Grant-in-Aid for JSPS Fellows, and Tsukuba Nanotechnology Human Resource Development Program, University of Tsukuba. K.I. and T.S. thank Professor Kiyoshi Asakawa of University of Tsukuba for his support. Sample patterning was done at the Nanofab facility of Institut N\'{e}el.

\bibliographystyle{apsrev}

\end{document}